# Modeling the optical properties of Twisted Bilayer Photonic Crystals

Haoning Tang[1], Fan Du[1], Stephen Carr[2], Clayton DeVault[1], Olivia Mello[1], and Eric Mazur[1*]

[1] School of Engineering and Applied Sciences, Harvard University, Cambridge, MA 02138, USA

[2] Brown Theoretical Physics Center and Department of Physics, Brown University, Providence, Rhode Island 02912, USA

[*] Email: mazur@seas.harvard.edu

**Abstract**

We demonstrate a photonic analog of twisted bilayer graphene that has ultra-flat photonic bands and exhibits extreme slow light behavior. Our twisted bilayer photonic device, which has an operating wavelength in C-band of the telecom window, uses two crystalline silicon photonic crystal slabs separated by a methyl methacrylate tunneling layer. We numerically determine the magic angle using a finite-element method and the corresponding photonic band structure, which exhibits a flat band over the entire Brillouin zone. This flat band causes the group velocity to approach zero and introduces light localization in a linear periodic photonic system. Using a plane-wave continuum model we find that photonic modes in twisted bilayer photonic crystals are not as tightly bound as their electronic counterparts in twisted bilayer graphene. In addition, the photonic system has a larger band asymmetry. The band structure can easily be engineered by adjusting the device geometry, giving significant freedom in the design of devices. Our work provides a fundamental understanding of the photonic properties of twisted bilayer photonic crystals and opens the door to nanoscale-based non-linear enhancement and superradiance.

**Introduction**

Over the past decade the stacking and twisting of two-dimensional materials has led to the development of novel materials with remarkable electronic properties. For example, in twisted bilayer graphene (TBG), an engineered material consisting of two stacked layers of graphene that are rotated relative to each other, at the so-called magic angle of $\theta = 1.1°$, the Fermi velocity drops to zero and the energy bands near the Fermi energy become flat [1]. These flat bands have high effective mass and half-filled correlated insulating states, resulting in superconductivity due to the formation of Moiré superlattices and Dirac cone hybridization [2, 3]. Exploring these unusual phenomena is central in the developing field of quantum twistronics [4, 5]. The concept of twistronics has been extended to include the study of nano-light properties in materials like TBG and twisted $\alpha$-MoO$_3$ [6-9]. Recently it was shown applying the ideas of twistronics to photonic Moiré lattices in dielectric nanophotonic materials lead to light localization and delocalization phenomena [10]. However, the connection between atomic twistronics and its nanoscale photonic analog has not been thoroughly explored.

Many concepts in condensed matter theory have photonic analogs. For example, photonic systems with nontrivial topological invariants are the photonic analog of the anomalous quantum hall effect and the anomalous quantum spin hall effect [11-19]. The periodic dielectric lattice of honeycomb lattice photonic crystals with "artificial atoms" (the unit cell in the dielectric structure) is analogous to the hexagonal atomic lattice of graphene. Indeed, these materials have been shown to give rise to topological photonics

[20-26]. In this context, it is natural to expect two layers of twisted honeycomb photonic crystal slabs to have similar physics as TBG. Yet while the acoustic analog to TBG has recently been demonstrated through phononic crystals [27], and while tunable light properties have been observed in metamaterials with Moiré patterns [28-31], a photonic band structure similar to the band structure in TBG-like systems has not been reported. Here, we correct that deficit.

In this paper we report on the modeling of twisted bilayer photonic crystals (TBPhCs) consisting entirely of dielectric materials. We find that TBPhCs have a photonic band structure that is similar to the electronic band structure of TBG. At a twist angle of $1.89°$, the resulting Moiré flat bands have group velocities ($v_g$) that vanish at the $K$ point leading to an extreme slow light effect. In analogy to the confinement of electronic wavefunctions in magic angle TBG, we observe low-loss light localization in this linear periodic photonic system. Unlike Anderson localization in optical quasicrystals, the localization we observe does not require disorder [10, 32]. Key differences between TBPhCs and TBG are that photonic states are not as tightly bound as their electronic counterparts and that the photonic system has a larger band asymmetry. The tunneling layer between the PhC slabs and the geometry of the slabs provide additional degrees of freedom for engineering the photonic band structure.

A major advantage of TBPhC over conventional slow-light media is that TBPhCs display slow-light behavior over an extremely narrow bandwidth. We can therefore design versatile TBPhCs that operate across a broad range of visible and infrared frequencies, which can be used to realize slow-light and flat band applications. These TBPhCs open the door to studying strong light-matter interactions, such as nonlinear enhancement and superradiance, where a combination of light localization, low loss, and slow light is required [33-36]. In addition, they can be used to investigate flat-band phenomena and wave-packet localization in two-dimensional systems at the nanoscale. Finally, the flexibility in designing TBPhCs permits simulating and exploring the band structure behavior of their electronic counterparts.

**Results**

Here we introduce a dielectric photonic crystal platform that hosts a band structure analogous to TBG. We start with a monolayer 2D honeycomb photonic crystal inspired by graphene [25]. The 2D photonic crystal is a silicon membrane with $C_{6v}$ symmetry-protected triangular shape air holes (Fig.1 a). We use a finite element method (COMSOL Multiphysics) to numerically calculate the band structure. The lowest singly degenerate quasi-transverse-electric (quasi-TE) band is well isolated from other higher-order bands. The $C_{6v}$ symmetry of the lattice also protects a Dirac-like crossing at the $K$ point centered at the Dirac cone frequency ($f_{DC}$), which is equivalent to the Fermi level in graphene (see Fig.1 b). TE-like electromagnetic modes that primarily propagate through air holes are still weakly coupled with neighboring holes, mimicking how electrons hop between carbon atoms in graphene.

In the honeycomb PhC, the nearest- and next-nearest-neighbor coupling strength can be controlled independently by varying the monolayer geometry, providing a platform to implement a broad class of tight-binding models. Building off this monolayer band structure, two sheets of photonic crystals are then coupled by an interlayer tunneling membrane to accurately recreate the AB- and AA-stacked configurations of bilayer graphene. In the AA-stacked configuration, two layers of PhCs are exactly aligned, while in the AB-stacked configuration, the top layer honeycomb center lies over one of the bottom layer's triangular airhole centers. The band structure of the AA-stacked configuration looks like two copies of the monolayer bands with a vertical offset of the Dirac cones at the $K$ point (see Fig.1 c). The AB-stacked configuration has a pair of touching parabolic bands with additional parabolic bands away from the touching bands (see Fig.1 d). Note that the AB- and BA-stacked configurations give identical band structures but not identical eigenmodes. The frequency separation between the bands in both stacking configurations is controlled by the tunneling strength between the two PhC layers, which is set by properties of both tunneling membranes and the PhC layers.

Next, we consider two adjacent PhC layers twisted by an angle $\theta$ relative to one another. This produces a Moiré pattern with a macroscopic periodicity of distinct AA and AB/BA stacking regions that grows in size as the angle decreases (Fig.1 e). Since our FEM calculation relies on the existence of Bloch waves (see S.1), we ensure that the structures created by twisting two lattices relative to each other are exactly periodic, or commensurate, by considering only specific twist angles [37],

$$\theta = 2\ arcsin\ arcsin\left(\frac{1}{2\sqrt{3n^2 + 3n + 1}}\right) \qquad \forall\ n \in \mathbb{Z}^+ \quad (1)$$

The twist angle controls the energy scale ($E = hf$) at which the Dirac cones of the two PhC layers intersect in momentum space. When this energy scale is comparable to the interlayer tunneling strength, band hybridization induces Moiré flat bands (see S.2). Those Moiré flat bands are fully compressed around the $f_{DC}$ and degenerate at the super lattice $K$ point (see Fig. 1 g-h). Our TBPhCs therefore reproduce similar band-flattening mechanism as TBG, eventually becoming flat with a zero $K$ point group velocity ($v_g(K) = 0$) at a "magic angle" of 1.89°.

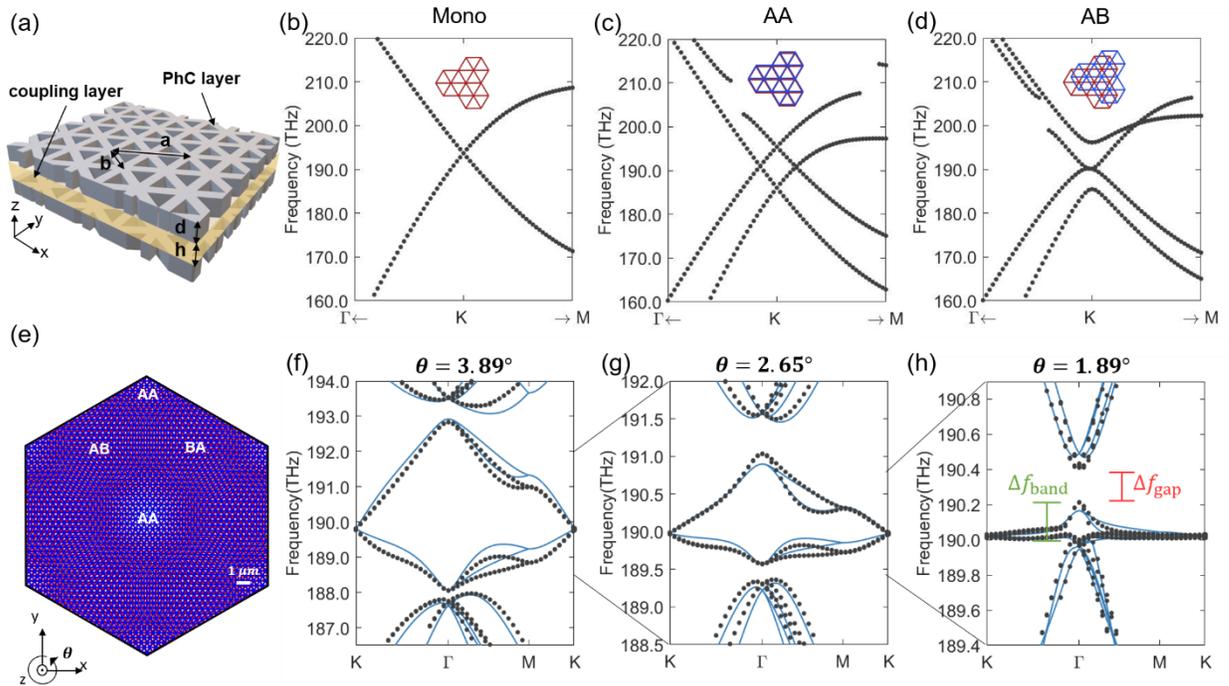

Figure 1  Twisted bilayer Photonic Crystals (a) Bilayer photonic crystal (BPhC) consisting of a tunneling layer with low refractive index sandwiched between two twisted dielectric layers. The 2D photonic crystal is a $d = 220\ nm$ thick silicon membrane ($n_{Si} = 3.48$) with $C_{6v}$ symmetry-protected triangular shape air holes. The triangular holes have a side length of $b = 279\ nm$ and the unit cell pitch is $a = 478\ nm$. The interlayer tunneling membrane has a thickness of $h = 250\ nm$ and refractive index of Polymethyl

Metacrylate (PMMA) $n_{PMMA} = 1.48$. (b-d) Simulated band structure of the monolayer (b), AA-stacked (c) and AB-stacked BPhCs (d). In (b-d) the insets show the respective real space configuration of the crystal unit cells. (e) Moiré pattern for a TBPhC, where the two dielectric layers are rotated by angle $\theta$ with respect to each other around the AA-stacked center. In a Moiré pattern, lattice structure locally resembles the regular stacking arrangement such as AA, AB, and BA. FEM band structure (Black dots) and fitted continuum model band structure (blue line) of (f) $\theta = 3.89°$. (g) $\theta = 2.64°$. (h) $\theta = 1.89°$. At small angles, the Dirac cones from each layer are pushed together and hybridize due to the interlayer tunneling.

Quasi-TE modes in the Moiré bands have symmetry properties and spatial profiles that agree with electronic wavefunctions in magic-angle twisted bilayer graphene [38]. For large angles, the Moiré quasi-TE modes are located across the entire supercell (Fig.2 a-b). At the magic angle $\theta = 1.89°$, as $v_g(K)$ vanishes, the quasi-TE modes become localized around the AA site (Fig.2 c). This type of localization is observed over most of the Brillouin zone except at the $\Gamma$ point where the AA site has zero mode intensity due to the symmetry (See S.3). All Moiré modes, including non-flattened Moiré modes, are all low-loss modes because their quality factors ($Q$-factor) vary from $2 \times 10^5$ to $3 \times 10^7$, while the modes in monolayer or AA/AB stacked photonic crystals have infinite $Q$-factor (See S.4). Although the Moiré modes have a similar mode profile as the monolayer eigenmodes, they do not maintain the translational symmetry of the monolayer eigenmodes because of a gradual change in intensity from AA site to AB/BA stacking sites. The mode symmetry mismatch explains our observation of slightly lower $Q$-factor in Moiré modes compared to monolayer and AA/AB stacked photonic crystal modes. The localization and high $Q$-factor properties of the Moiré modes could be beneficial in the realization of nonlinear enhancement and superradiance when combined with the slow-light effect we investigated in the next section.

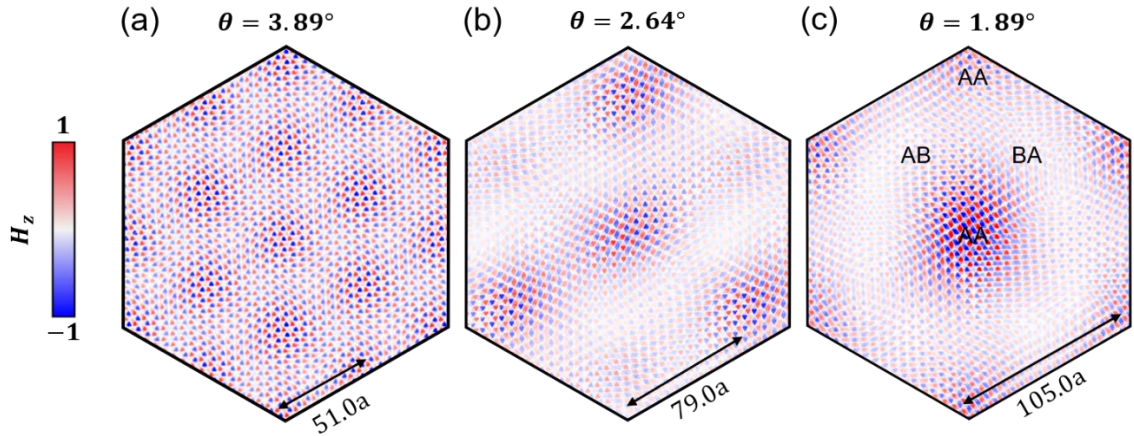

Figure 2    (a) For large angles 3.89°, the TE-like modes at the $K$ points are itinerant and persist across the entire supercell. (b) At 2.64° they become localized on the AA-stacked region in the center of the supercell as $\theta$ decreases. (c) The quasi-TE modes are mostly localized when $\theta = 1.89°$.

Compared to TBG's electronic band structure, the TBPhCs Moiré bands have a stronger twist dependence on the Dirac cone frequency $f_{DC}$, which moves $0.2\ THz$ (0.51% of its' first nearest-neighbor coupling amplitude, which in TBG would correspond to roughly 15 meV variation in the energy of the Dirac cones) between 3.48° and 1.89°, and more severe top-bottom band asymmetry for both the Moiré band gaps ($\Delta f_{gap}$) and $K$ point group velocities ($v_g(K)$). Compared to graphene, the band gaps ($\Delta f_{gap}$) at the top ($f > f_{DC}$) and bottom ($f < f_{DC}$) bands have different frequency scales, with the top gap $3 - 4$ times larger than the bottom gap (See Fig.3 (a)). We also observe that at angles larger than 3°, the bottom bands show a much slower dispersion than the top bands away from the Dirac frequency.

We investigate the origins of these differences and try to reproduce these features by considering a low-energy expansion of the TBG's band structure. The effective Hamiltonian consists of Dirac Hamiltonians from both layers, sampled on momenta which are scattered by the Moiré reciprocal lattice, and off-diagonal inter-layer tunneling terms [5]. We begin with the block diagonal part, which are the Dirac Hamiltonians given in terms of the relative momentum $q$ away from a $K$ point ($O(q^2)$):

$$H_{gr}(K+q) \approx \frac{-a(t_1-2t_3)\sqrt{3}}{2}\sigma \cdot q - \frac{3a^2 t_2}{4}q^2 - 3t_2 \qquad (2)$$

where $\sigma$ are the $2 \times 2$ Pauli matrices and $a$ is the lattice constant. Some of the unconventional behavior can already be explained by this monolayer Hamiltonian. If the effective second nearest-neighbor coupling term ($t_2$) changes by a fraction of a $THz$ between different angles, due to relatively strong tunneling between the photonic states of the two layers, then the Dirac cone frequency $f_{DC}$ is affected by the last term in Eq. 2. A large value of $t_2$ also explains the difference between the top and bottom $v_g(K)$ at large angles, due to the frequency-asymmetric $q^2$ term.

We now move to the block off-diagonal terms in the effective Hamiltonian. The interlayer tunneling in TBG between pairs of orbital types of different layers (say, AA or AB) varies smoothly with the periodicity of the Moiré superlattice. This justifies their parameterization by just their first-order Fourier coefficients: commonly labeled $\omega_0$ for tunneling between orbitals of the same type (AA and BB) and $\omega_1$ for orbitals of differing types (AB and BA) [1]. To open up significant superlattice gaps, $\omega_0$ must be smaller than $\omega_1$, with $\omega_0 = 0$ maximizing the superlattice gap, while $\omega_1$ defines the effective tunneling strength for the magic-angle condition [39]. This simple model produces bandgaps of similar size for both the top and bottom bands. However, a large disparity in the size of the top and bottom gaps has been seen before in DFT calculations of Lithium intercalated TBG [40]. In that case, it was argued that the effective interlayer tunneling strength for the top(bottom) bands were very different, due to the Li atoms' charge transfer preferentially enhancing or screening the tunneling between the layers at different energies. As the photonic crystal states are not as tightly bound as the $p_z$ orbitals in graphene (see our later estimations of the monolayer $t_i$ values), having an asymmetry in the effective interlayer tunneling at high and low energies is even less surprising here. Therefore, we fit our continuum model to the TBPhCs FEM band structures by tuning variables in the following manner: for the monolayer model, pick fixed values of $t_1$, $t_2$, and $t_3$ across all twist angles; for the interlayer tunneling, pick $\omega_0$ and $\omega_1$ independently for the top and bottom bands, giving four variables: $\omega_0^t, \omega_1^t, \omega_0^b, \omega_1^b$. In addition, near the magic-angle these terms should become similar, so we allow them to generically depend on $\theta$.

We find that $[t_1, t_2, t_3] = [-39, 17, -5]\ THz$ works well for all twist angles. For the low energy Hamiltonian, only the term $t_1 - 2t_3$ enters as the $v_g(K)$, so we can freely increase $t_1$ if we also increase $t_3$ by twice that. The asymmetry in the $v_g(K)$ sets the strength of $t_2$, and so we increase $t_1$ and $t_3$ together to give a sequence of couplings that show reasonable decay in strength. In contrast, for TBG the coupling strengths given by DFT simulations decay by roughly a factor of 10 between $t_1$ and $t_2$ [41], indicating that these photonic states are not as tightly bound as their electronic counterparts.

At the magic angle ($\theta = 1.89°$) with the default photonic crystal parameters, we find good agreement when selecting $\omega_0^t = 1.43$ and $\omega_1^t = 1.85\ THz$ for the top-side tunneling and $\omega_0^t = \omega_1^b = 1.85\ THz$ for the bottom-side tunneling. When the tunneling strengths depend linearly on the twist angle the bands are well captured. This linear dependence is found to be

$$\omega_i^t(\theta) = \big(1 - 0.15(\theta - \theta_m)\big)\omega_i^t(\theta_m) \qquad \omega_i^b(\theta) = \big(1 + 0.15(\theta - \theta_m)\big)\omega_i^b(\theta_m) \qquad (3)$$

for $\theta$ evaluated in degrees. At $\theta = 4°$, the tunneling coefficients are roughly 30% weaker for the top-side and 30% stronger for the bottom-side. The enhancement of the bottom-side tunneling is counterbalanced by a suppression of the top-side tunneling, implying that the "net" tunneling is unchanged while its distribution between the relevant orbitals is modified, in agreement with the study of Li-intercalated TBG. This unusual interlayer tunneling behavior results in the severe top band and bottom band asymmetry we observe in the TBPhCs band structure.

Our model gives a very reliable reproduction of the bands close to the Dirac cone frequency, with small disagreements only occurring in the top-bottom band symmetry of the flat bands right at the magic angle tuning. This is due to terms not captured by our simple model. Specifically, we omit the momentum-dependence of the interlayer tunneling terms [42, 43] that provide a more accurate description of their Fourier transform.

Photonic Moiré flat bands at both a magic-angle $\theta = 1.89°$ and $\theta = 2°$ exhibit slow-light effects. Here we show the group velocity ($v_g = d\omega/dk$) was drastically reduced in the $\Gamma$ to $K$ direction. (See Fig.3 (b)). Compared to the light propagation in the monolayer PhC, at $\theta = 2°$, $v_g$ was reduced from $0.2c$ to $0.005c$ near the $K$ point and from $0.65c$ to $0.007c$ near the $\Gamma$ point. Similarly, at the magic angle $\theta = 1.89°$, $v_g$ was reduced to zero at $K$ point and to $0.01c$ near $\Gamma$ point. Note that although the magic angle is at $\theta = 1.89°$, the TBPhCs at $\theta = 2°$ has narrowest Moiré bandwidth ($\Delta f_{band}$) of $0.217\ THz$ (See Fig.2 (a)). This is because of a higher $v_g$ around $\Gamma$ point at $1.89°$. Notice that we have small $v_g$ band over the entire Brillouin zone instead of only on the band edge, which is a crucial for all-directional photonic devices. This unconventional slow light effect and narrow bandwidth indicate great opportunity in nonlinear enhancement and superradiance.

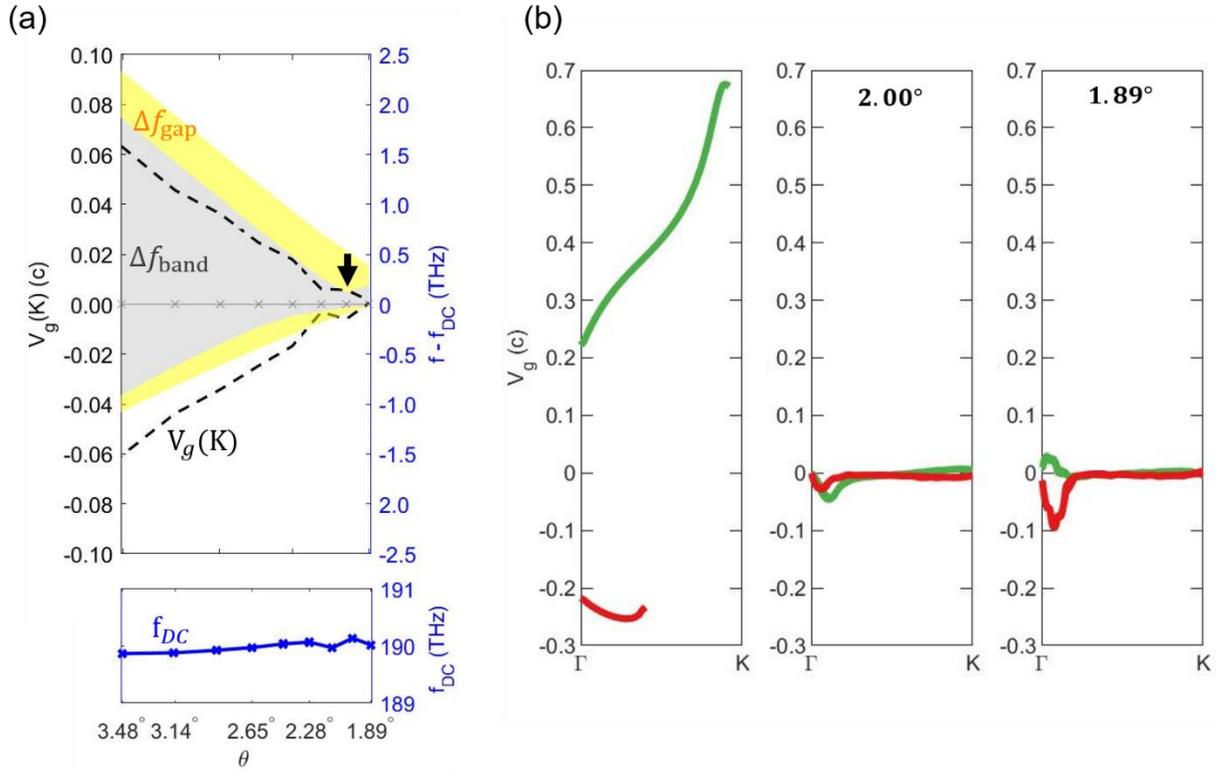

Figure 3 (a) Top and bottom band gap $\Delta f_{gap}$ (yellow), bandwidth $\Delta f_{band}$ (gray), $v_g(K)$ (black dashed line), and Dirac cone frequency $f_{DC}$ for small commensurate angles. The narrowest bandwidth is pointed by the black arrow. (b) Group velocity of monolayer configuration (gray) and the magic angle configuration, the line discontinuity is due to the light-cone, upper part of Moiré bands (green) and bottom part of Moiré bands (red) has different group velocity from $\Gamma$ to $K$.

We can further study the band structure behavior given the flexibility of tuning the geometry parameters in TBPhCs. We fix $\theta = 2.13°$ and change the tunneling layer thickness ($h$), the refractive index ($n_{tunneling}$) of the tunneling layer, and the refractive index of PhC bilayer individually ($n_{PhC}$). The Dirac cone frequency is modified when tuning $h, n_{tunneling}$, and $n_{PhC}$ due to the change of the effective refractive index of TBPhCs. Band structures have different dependency on ($h, n_{tunneling}, n_{PhC}$) because the interlayer tunneling was modified in different ways. When decreasing $h$, $f_{DC}$ remains unchanged. We obtained the lowest value of $V_g(K)$ at $h = 250\ nm$ and the narrowest bandwidth $\Delta f_{band} = 0.185\ THz$ at $h = 240\ nm$ (see Fig.4 (a)). By increasing $n_{tunneling}$, we can decrease $f_{DC}$ at the same time. $v_g(K)$ was reduced to zero when $n_{polymer} = 1.59$, while the narrowest bandwidth $\Delta f_{band} = 0.18\ THz$ is obtained at $n_{tunneling} = 1.55$ (see Fig.4 (b)). By decreasing $n_{Si}$, we increase $f_{DC}$, and $v_g(K)$ was reduced to zero when $n_{Si} = 3.05$, the narrowest bandwidth $\Delta f_{band} = 0.2\ THz$ is obtained at $n_{Si}=3.3$ (see Fig.4 (c)). We again fit our continuum model to the TBPhCs FEM band structures and study the interlayer tunneling dependency on these three geometry parameters (See S.5). For these three photonic geometry variables, we start with the default setting $h = 250nm, n_{tunneling} = 1.48$, and $n_{PhC} = 3.48, h = 250nm$, and obtain the following dependence in the tunneling parameters in the plane-wave continuum model,

$$\omega_1^b = \omega_1^b(\theta_m)$$

$$\omega_i^j = \beta \omega_i^j(\theta_m), \text{ else} \quad (4)$$

where $\beta$ is a scaling factor dependent on the selected geometry variable, following:

$$\beta(h) = 1 - \frac{h - 250\ nm}{135\ nm}$$

$$\beta(n_{PhC}) = 1 - 0.85(n_{PhC} - 3.48)$$

$$\beta(n_{tunneling}) = 1 + (n_{tunneling} - 1.5) \quad (5)$$

The dependency in tunneling parameters matches the direct observation from the FEM band structure: changing the geometry parameters tends to affect all tunneling parameters except the AA orbital tunneling represented by $\omega_1^b$. As $h$ increases, $n_{tunneling}$ increases, or $n_{PhC}$ decreases, the strength of the tunneling becomes smaller. It is possible that the tuning modifies not only the $\omega$ terms but also the in-plane couplings $t_i$, but near the magic-angle the band structure is predominantly defined by a ratio between these two types of model parameters [1], therefore we consider the $t_i$ fixed for simplicity. The parameter-independence of $\omega_1^b$ is motivated by observations of near parameter-independence of the bottom bands, and that tuning $\omega_1^b$ by $\beta$ reduces the agreement with FEM band structures. This is likely because the geometry parameters predominantly modify the higher frequency photonic modes, e.g., the top half bands (see S.6). Knowing the geometry dependency gives more freedom in further engineering the optical Moiré flat band and indicate the possibility of having magic angle at higher $\theta$, where localized modes are closer to each other in distance [27].

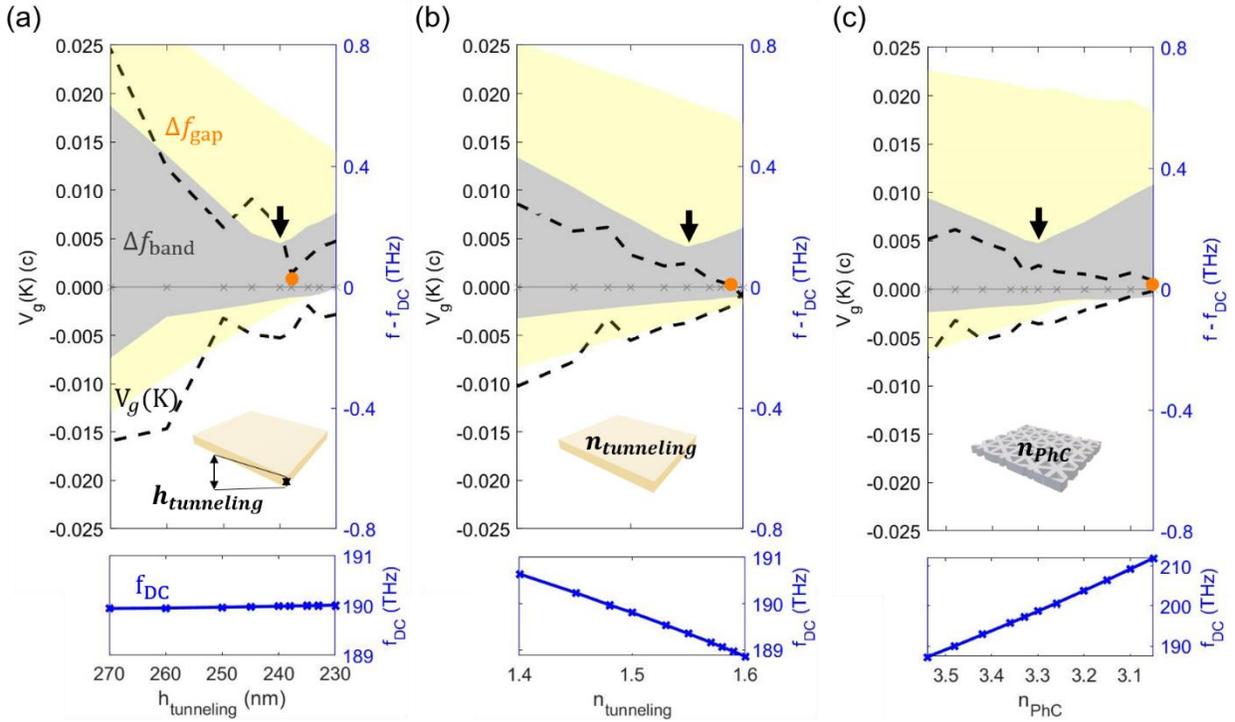

Figure 4 (a) Top and bottom band gap $\Delta f_{gap}$ (yellow), bandwidth $\Delta f_{band}$ (gray), group Velocity at $K$ point $v_g(K)$ (black dashed line), and Dirac cone frequency $f_{DC}$ when changing (a) $h$ (b) $n_{tunneling}$ (c)

$n_{PhC}$. The narrowest bandwidth is pointed by the black arrow and the vanishing $v_g(K)$ is represented by the orange dot.

**Discussion**

Our numerical calculations show that the dispersion of electromagnetic waves can be manipulated dramatically, from highly dispersive to flat, by simply changing the angle between two photonic crystal slabs. We identified the magic angle where Moiré flat bands appear, leading to a 40–120-fold reduction in group velocity compared to monolayer PhC. At this angle TBPhCs exhibit slow-light behavior within an extremely narrow bandwidth and the eigenmodes are highly localized in the regions exhibiting AA stacking. We studied the photonic band structure behavior using a plane-wave continuum model and found that TBPhCs differ from TBG both in intralayer coupling and interlayer tunneling characteristics. We find that interlayer tunneling can be controlled by tuning the geometry parameters ($h$, $n_{tunneling}$, $n_{PhC}$), facilitating the design of an optical flat band.

The "twisted photonic crystal toolkit" we present here provides access to slow-light effects and light localization that cannot be accomplished by conventional photonic crystals. Therefore, TBPhCs will drastically enhance access to optical nonlinearities and quantum interactions in photonic devices. Because TBPhCs are designed for standard silicon-on-substrate wafers and can be fabricated by a wafer bonding and transferring technique, the fabrication of such devices is immediately feasible.

**Materials and Methods**

Simulation: The FEM band structure, eigenmode, and $Q$- factor simulations were computed using three-dimensional finite element methods (COMSOL Multiphysics 5.4). We first calculated all the modes in a PhC unit cell/super unit cell with Floquet periodic boundary conditions in the two lattice-vector directions and perfectly matched layers at the boundaries in the out-of-plane direction. TM/TE-polarized modes were selected by evaluating the energy ratio of the electric and magnetic fields in all directions. The simulations were carried out on a Dell M630 computer (2 x Intel Xeon CPU E5-2697 v4 2.30GHz 18 core, 247GB RAM, 1GbE, FDR Infiniband). The time to calculate the photonic band structure at the magic angle is roughly 24 hours. The plane-wave continuum model is implemented using MATLAB. The estimated calculation time is a few seconds per band structure.

**Author Contributions**

H.T and C.D. provided the initial idea for this work. H.T., D.F., and O.M. carried out the FEM simulations. S.C. carried out the continuum model fitting and analysis. E.M. supervised the research and the development of the manuscript. H.T. wrote the manuscript with input from all authors. All authors subsequently took part in the revision process and approved the final copy of the manuscript.

**Notes**

The authors declare no competing financial interest.

**Acknowledgment**

The authors thank Jennifer E. Hoffman, Minhal Gardezi, Harris Pirie, William Dorrell, S. Ben November, Fanqi Yuan, Shang Liu, Renjing Xu, Carlos Rio, Michaël Lobet, Zhu Yuan and Weilu Shen for discussions. The Harvard University team acknowledges support from DARPA under contract URFAO: GR510802. The FEM simulations in this paper were run on the FASRC Cannon cluster supported by the FAS Division of Science Research Computing Group at Harvard University.

**Supplementary Information for Modeling the optical properties of Twisted Bilayer Photonic Crystals**

Haoning Tang[1], Fan Du[1], Stephen Carr[2], Clayton DeVault[1], Olivia Mello[1], and Eric Mazur[1*]

[1] School of Engineering and Applied Sciences, Harvard University, Cambridge, MA 02138, USA

[2] Brown Theoretical Physics Center and Department of Physics, Brown University, Providence, Rhode Island 02912, USA

[*] Email: mazur@seas.harvard.edu

A simplified continuum model of bilayer graphene can be understood by beginning with a nearest-neighbor tight-binding model. This model consists of two Carbon atoms of the honeycomb lattice of graphene, with lattice vectors $a_1 = a\left(\frac{\sqrt{3}}{2}, -\frac{1}{2}\right), a_2 = a\left(\frac{\sqrt{3}}{2}, \frac{1}{2}\right)$ and reciprocal vectors $b_1 = \frac{2\pi}{a}\left(\frac{1}{\sqrt{3}}, -1\right), b_2 = \frac{2\pi}{a}\left(\frac{1}{\sqrt{3}}, 1\right)$. The $A$ and $B$ sublattice atoms are positioned at 0 and $\frac{1}{3}(a_1 + a_2)$, respectively. Assuming a nearest neighbor coupling $t_1$, a second-nearest neighbor coupling of $t_2$, and a third-nearest neighbor coupling of $t_3$, the low-energy Hamiltonian for a momentum $K + q$ near the Dirac cone $K = \frac{b_1 - b_2}{3} = \frac{2\pi}{a}(0, -2/3)$ is, up to $(O(q^2))$:

$$H_{gr}(K+q) \approx \frac{-a(t_1 - 2t_3)\sqrt{3}}{2} \sigma \cdot q - \frac{3a^2 t_2}{4} q^2 - 3t_2 \qquad (5)$$

where $\sigma$ are the $2 \times 2$ Pauli matrices, taken as a rank 3 tensor acting on the momentum vector $q$. The group velocity at $K$ (slope of the bands at the cone) is $V_g(K) = \frac{a(t_1 - 2t_3)\sqrt{3}}{2}$. There is also a top-bottom band symmetry-breaking term of strength $\frac{3a^2 t_2}{4}$, which goes like $q^2$ along the diagonal of $H$, causing both top and bottom levels away from $K$ point to be pushed in the same direction.

The interlayer coupling between two graphene monolayers can also be highly simplified.
The coupling between any pair of interlayer orbital types can be represented with a first-order Fourier expansion, consisting of three plane-waves of equal strength summed together, and with the relative phases of each term chosen appropriately such that the maxima are centered at the appropriate location in the Moiré superlattice. This introduces $2 \times 2$ scattering matrices with specific phases for the three-fold related scattering momenta, namely,

$$T_1 = (\omega_0\ \omega_1\ \omega_1\ \omega_0\ ), T_2 = (\omega_0 \psi^*\ \omega_1\ \omega_1 \psi\ \omega_0 \psi^*), T_3 = T_2^* \qquad (6)$$

for $\psi = e^{i2\pi/3}$.

The bilayer Hamiltonian can then be easily formulated by combining these two simplifications:

$$H(q) = (H_D(q)\ \Sigma T_i\ \Sigma T_i^\dagger\ H_D(q)) \qquad (7)$$

where $q$ is understood to be taken relative to the $K$ point of monolayer graphene's Brillouin zone, and the $T_i$ scatter between different $q$ by reciprocal vectors of the Moiré superlattice ($q_1 = 0$, $q_2 = G_2$, and $q_3 = -G_1 + G_2$, for $G_i$ the reciprocal vectors of the Moiré supercell).

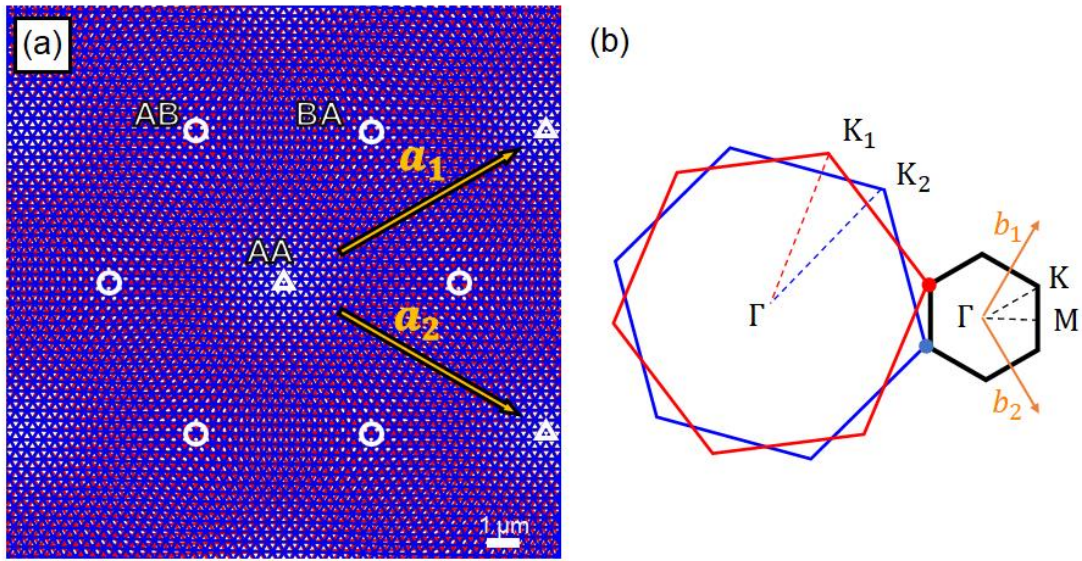

S.1 (a) Photonic crystal structure and Moiré pattern of TBPhCs with $\theta = 1.89°$, the top(bottom) layer is blue(red). AA and AB (or BA), are indicated by triangles and circles, respectively. The triangular superlattice vectors $a_1$ and $a_2$ are triangular superlattice vectors. (b)The super lattice Brillouin zone is constructed from the wave vectors of the two photonic crystals layers. and the small hexagon is the Moiré Brillouin zone of TBPhCs with reciprocal vectors $b_1$ and $b_2$.

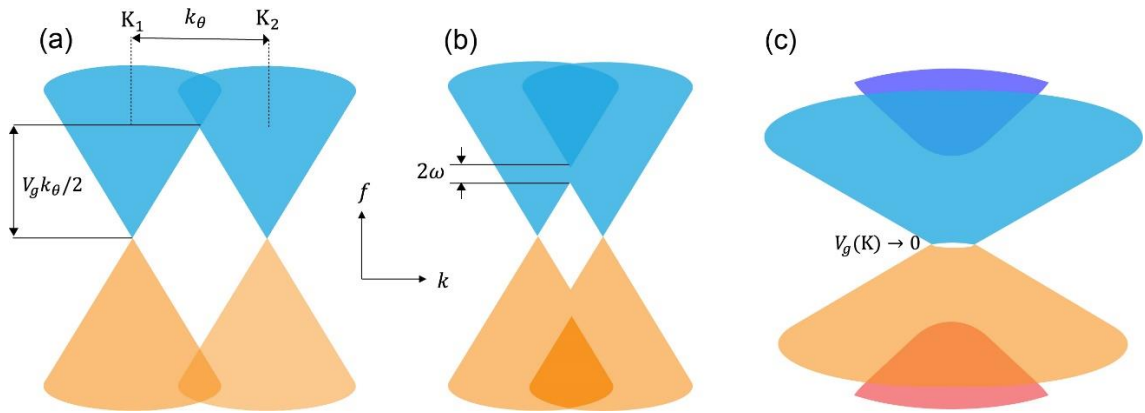

S.2 (a) Rotation of the Brillouin zones causes the Dirac cones of the two layers to rotate and intersect with each other. The hybridization strength $\omega = 0$ (b) When there is enough electron tunneling between two layers, those Dirac-cone bands further band hybridize with each other. $2\omega \ll v_g k_\theta / 2$ (c)Finally, dirac-cone bands get closer to each other and it becomes flat and $2\omega \sim v_g k_\theta / 2$. In the middle between red and blue. The flattening of the bands near the magic angle can be understood from the competition between the kinetic energy and the interlayer hybridization energy.

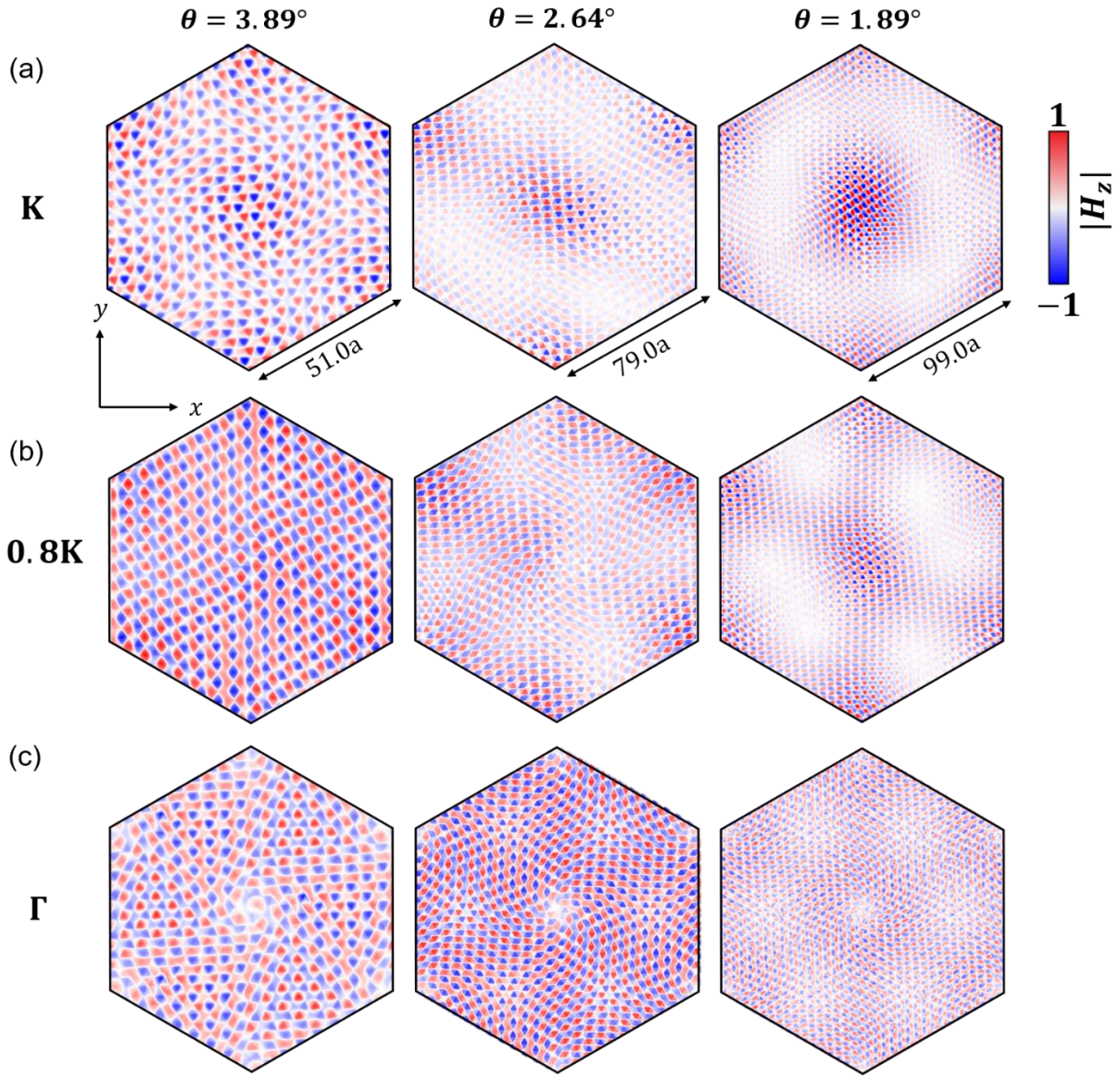

S.3  Eigenmode localization at (a)$K$ (b)$0.8K$ and (c) $\Gamma$ point. At small angles, the modes are localized around the AA stacked region. Due to symmetry, the AA site is zero for $\Gamma$ wavefunction.

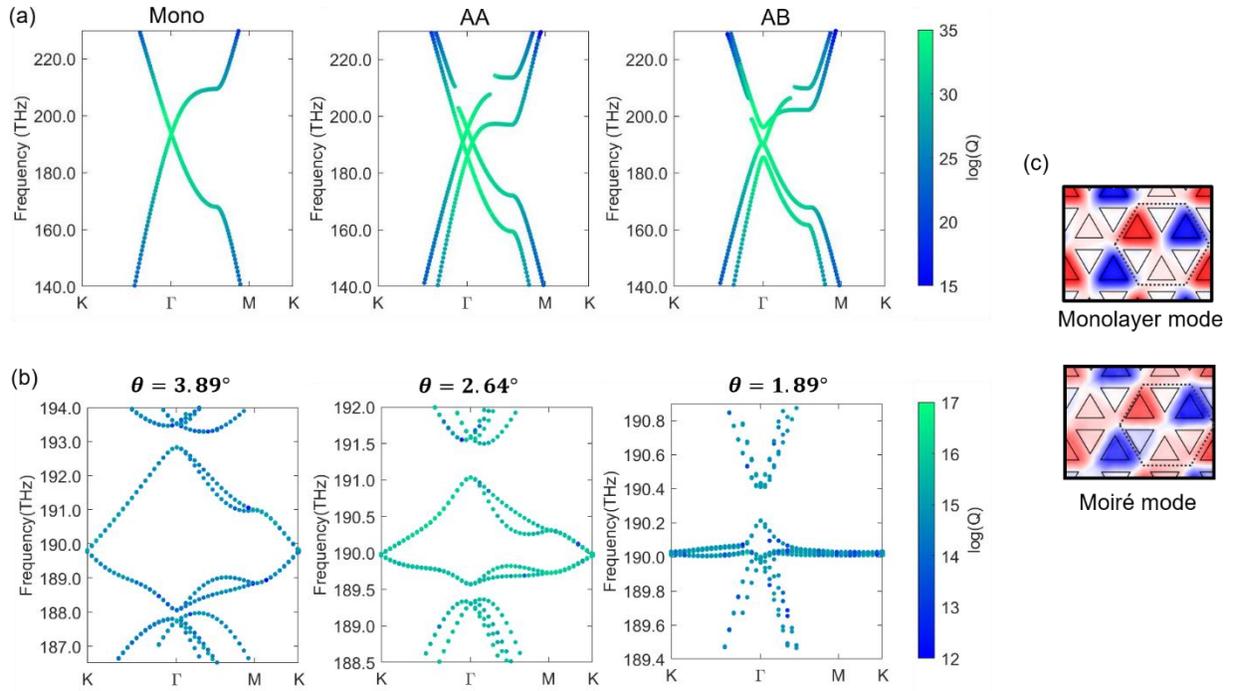

S.4  (a)The quality factors ($Q$-factor) of the monolayer PhC, AA-stacked PhCs and AB-stacked PhCs are very high over the entire Brillouin zone and is infinite near $K$ point. (b) The quality factors ($Q$-factor) of the TBPhCs are high ($2 \times 10^5$ to $3 \times 10^7$) over the entire Brillouin zone but not infinite. (c) Monolayer eigenmode at $K$ point and Moiré eigenmode at $K$ point has the same mode configuration but different symmetry due to the different intensities at triangular airhole sites. The magnitude of the Moiré mode in one hexagonal cluster (dash line) has a gradual change from AA to AB/BA stacked region. This explains why some of the Moiré eigenmode leaks to the air and results in a lower $Q$-factor.

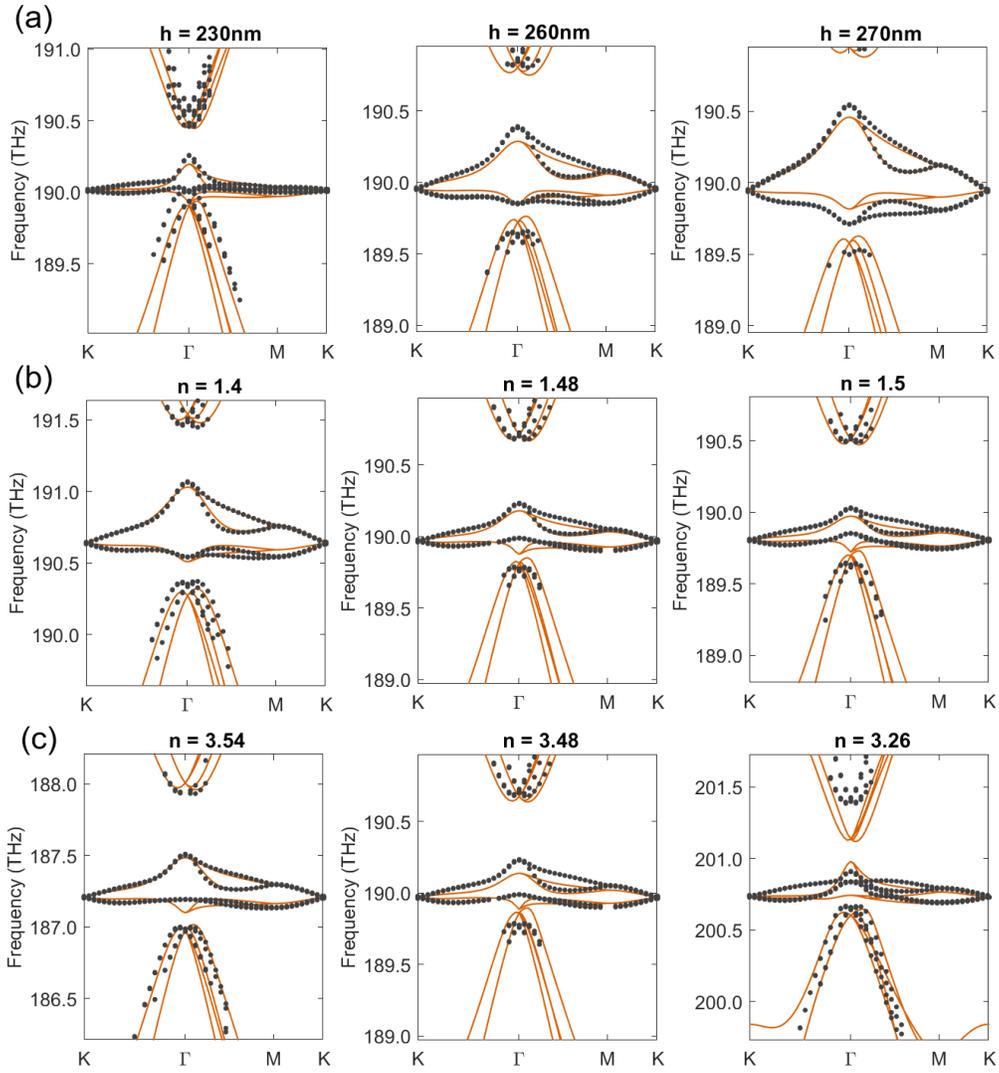

S.5 (a) Fitting of the continuum model (lines) to the FEM results (black dots) for changes in (a) $h$ (b) $n_{tunneling}$, and (c) $n_{PhC}$

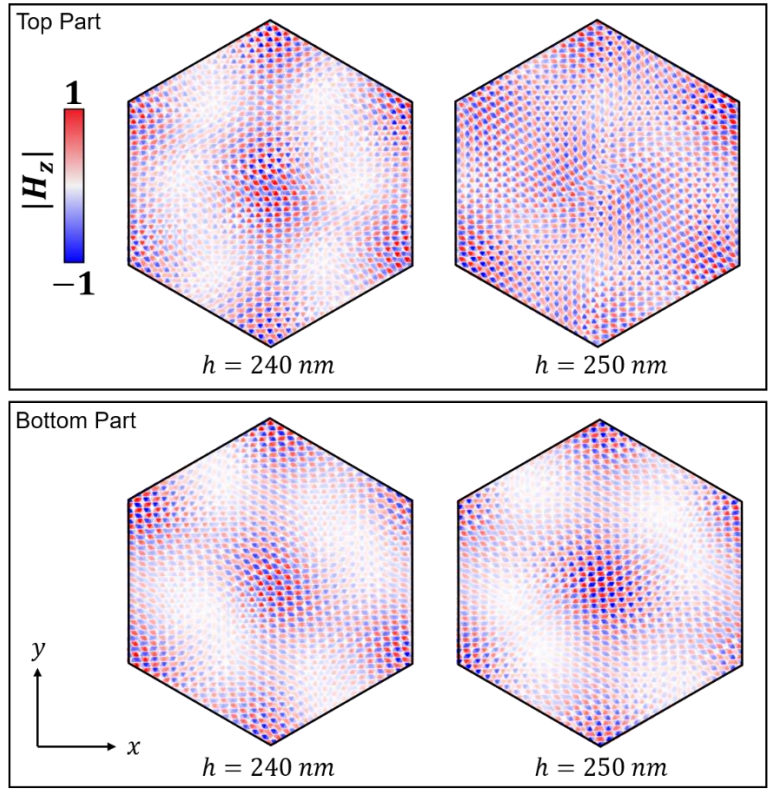

S.6     Geometry parameters like the thickness of the tunneling layer mostly modify the higher frequency photonic modes in the top part of Moiré bands. Changing the geometry parameters does not modify the mode configuration of the bottom part of the Moiré bands very much.